\documentclass[10pt]{article}
\usepackage[letterpaper]{geometry}
\usepackage{hicss}
\usepackage{times}
\usepackage[none]{hyphenat}
\usepackage{url}
\usepackage{latexsym}
\usepackage{minted}
\usepackage{indentfirst}
\usepackage{graphicx}
\usepackage{booktabs}
\usepackage{multirow}
\usepackage{rotating}
\usepackage{threeparttable}
\graphicspath{{images/}}
\usepackage[
    style=apa,
  ]{biblatex}
\usepackage{tabularx} 
\usepackage{ragged2e} 

\title{Can AI Recognize Its Own Reflection? Self-Detection Performance of LLMs in Computing Education}

\author{Christopher Burger \\
 The University of Mississippi \\
 {\underline{ cburger@olemiss.edu}} \\ \\
 \\ \And
 Karmece Talley \\
 Rust College \\
 {\underline{ karmecet1@gmail.com} } \\ \\
 \\ \And
 Christina Trotter \\
 The University of Mississippi \\
 {\underline{ cjtrotte@olemiss.edu} } \\ \\
 }

\date{}

\begin{document}
\maketitle
\begin{abstract}

The rapid advancement of Large Language Models (LLMs) presents a significant challenge to academic integrity within computing education. As educators seek reliable detection methods, this paper evaluates the capacity of three prominent LLMs (GPT-4, Claude, and Gemini) to identify AI-generated text in computing-specific contexts. We test their performance under both standard and `deceptive' prompt conditions, where the models were instructed to evade detection. Our findings reveal a significant instability: while default AI-generated text was easily identified, all models struggled to correctly classify human-written work (with error rates up to 32\%). Furthermore, the models were highly susceptible to deceptive prompts, with Gemini's output completely fooling GPT-4. Given that simple prompt alterations significantly degrade detection efficacy, our results demonstrate that these LLMs are currently too unreliable for making high-stakes academic misconduct judgments.

\end{abstract}

\subsubsection*{Keywords:}

Large Language Models, Artificial Intelligence, Academic Integrity, Self-detection, Machine Learning Ethics

\section{Introduction}
The rapid advancement of Large Language Models (LLMs) has transformed how students engage with academic content, especially in computing education. These models can now generate coherent explanations of algorithms (or at least reasonable enough to expect that the errors made are the products of human thought), produce detailed documentation, and even write functional code with minimal human interaction. While offering valuable learning opportunities, this technological leap has simultaneously created significant challenges for academic integrity assessment. Computing educators increasingly face the difficult task of determining whether student submissions are original work or AI-generated content, a distinction critical to maintaining educational standards and ensuring authentic skill development. The integration, either explicitly or otherwise, of LLMs into university settings has brought forth urgent questions about detection capabilities and appropriate policies for addressing AI-assisted work in computer science and engineering education.

The importance of addressing LLM-generated content in computing education stems from the field's necessary requirement of establishing fundamental problem-solving skills, which can never be learned when students leverage AI tools indiscriminately. Unlike traditional plagiarism, which can often be identified through existing detection tools, LLM-generated content requires more effort to identify as by design it produces output tailored to specific prompts combined with inherent stochasticity that results in similar, but non-identical output between successive identical queries to the model. This distinction has left educators at universities struggling to establish consistent evidence-based policies, and there is substantial debate on what such a policy should even contain \parencite{zhou2024theteachersconfusedwell, Fowler100daysGPT}. Computing departments, in particular, find themselves at the center of this issue, as their students typically possess the baseline technical understanding needed to effectively use these tools (largely through writing effective prompts). Without reliable detection methods, educators risk either incorrectly penalizing innocent students or failing to identify genuine instances of academic dishonesty.

The detection of LLM-generated content presents considerable technical difficulties, especially given the rapid advancement in model quality over the last few years. Traditional computational linguistics approaches that once differentiated machine-generated content now struggle with sophisticated models that have been trained on vast corpora of human writing. This challenge is compounded with the inclusion of `personalities', where human-like mannerisms and idiosyncrasies can be set as a kind of filter that the generated content passes through to better suit a user's preference or for some particular purpose.. Furthermore, the technical nature of computing content often features specialized vocabulary and structured explanations that may naturally resemble LLM outputs even when human-authored. These factors collectively create a detection landscape where both false positives and false negatives can carry significant consequences for educational assessment and student outcomes.


Our research confronts these challenges through a systematic evaluation of three prominent LLMs (GPT-4, Claude, and Gemini) in computing-specific contexts. We designed an experiment to test their detection accuracy under both standard conditions and a deliberately `deceptive' mode, where models were instructed to evade detection. The results reveal a substantial flaw: while default AI-generated text was easily identified, the models' performance collapsed when evaluating deceptively generated content or even human-authored work. By demonstrating that simple prompt alterations are sufficient to fool these systems, our work provides important evidence that current LLMs lack the reliability required for high-stakes academic integrity decisions.
\section{Background and Related Work}
The field of Large Language Model (LLM) development and analysis is advancing at an exceptionally rapid pace. Consequently, much of the foundational and most current research is first made available through preprint archives like arXiv. While we prioritize citing peer-reviewed sources where possible, this work also references several key preprints. We adopt the approach used by leading organizations like the Association for Computational Linguistics which requires all relevant work regardless of review status be included as a measure of good faith to ensure a timely and relevant discussion of the state-of-the-art. We have critically evaluated these sources and believe their methodologies and findings are sound and directly relevant to our investigation.

The recent rapid development and adoption of LLMs has fundamentally transformed the landscape of academic work and educational assessment. The wide availability of LLMs has resulted in various reactions among both students and educators, but there is now acceptance among the latter that LLMs have sufficient capability to solve many fundamental entry-level problems. As such, LLM usage needs to be understood as likely outside of proctored environments since even institutions with restrictive policies still have substantial student LLM use \parencite{zhu2024embracingaieducationunderstanding}. Zhu et al. also indicate that 70\% of surveyed middle and high school students have utilized LLMs, an important sign that the trend of increasing LLM usage among students will continue.

In computing education specifically, the implications of LLM usage are especially significant. Students now have access to tools capable of generating code explanations, algorithm descriptions, technical documentation, and complete programming solutions with substantial accuracy while requiring only a modest level of interaction with the model. This has created what many educators describe as a paradigm shift in how academic integrity is conceptualized and enforced, particularly in disciplines where problem-solving processes and technical communication skills are core learning objectives.

\subsection{LLM Content Detection}
The field of LLM-generated text detection has emerged as a research area substantially in part due to industry concerns about AI generated code. Flaws \parencite{liu2023codegeneratedchatgptreally,mastropaolo2023robustnesscodegenerationtechniques}, security vulnerabilities \parencite{fu2025securityweaknessescopilotgeneratedcode,} and intellectual property infringement \parencite{pmlr-v202-yu23g} are significant concerns for both private developers and researchers who use LLM generated code. As such, AI code detection is now a significant field of interest with a growing background of literature \parencite{suh2024empiricalstudyautomaticallydetecting}. LLM-generated text detection attempts to identify if a piece of text was produced by a LLM, which is effectively a standard binary classification task. Current detection approaches can be broadly categorized into several methodologies:
\\

\noindent\textbf{Statistical Linguistic Approaches:} Traditional computational linguistics methods focus on identifying statistical patterns that differentiate human and machine-generated text. These approaches use machine learning to analyze text and look for the particular indicators with of AI-generated text. These indicators are with respect to some metric, like \textit{perplexity} (a measure of text complexity which is generally lower in human authored text) and \textit{burstiness} (a measure of sentence length uniformity, where humans tend to be less consistent in sentence length than LLMs). However, as LLMs become more sophisticated and better replicate human content, the deviations detected by these metrics become less substantial which ultimately results in less reliable detection.
\\

\noindent\textbf{Watermarking Techniques:} Watermarking is a proactive approach to LLM detection, where identifiable patterns are embedded during text generation. \textcite{pmlr-v202-kirchenbauer23a} demonstrated promising results with watermarking algorithms that can maintain text quality while enabling reliable detection. Unfortunately, these methods require implementation at the generation stage and cannot address content from non-watermarked models. As models that students would use possess no watermark, this approach is not currently a practical solution.
\\

\noindent\textbf{Zero-shot Detection Methods:} DetectGPT \parencite{mitchell2023detectgptzeroshotmachinegeneratedtext} and similar zero-shot approaches have shown some promise in controlled settings. These methods use the model's own output distribution to identify generated content without requiring training on specific model outputs. However, the reliability of these detectors in real-world applications remains suspect \parencite{10.1145/3639474.3640068, wu2025detectrlbenchmarkingllmgeneratedtext}. 
\subsubsection*{Self-Detection Capabilities of LLMs:}

A recent line of research examines whether LLMs can detect their own or other model's outputs.  \textcite{caiado2023aicontentselfdetectiontransformerbased} directly investigated the capacity for transformer-based LLMs to detect their own output. We are especially interested in how their findings relate to the improvements in LLMs over the approximately 1.5 years since their work was released. 
\textcite{li2024thinktwicetrustingselfdetection} constructed a framework to reduce over-certainty in LLM self-output evaluation. \textcite{wu2025detectrlbenchmarkingllmgeneratedtext} developed a benchmark, DetectRL, demonstrating that state of the art detection methods (some of which are `sibling' LLMs like RoBERTa )  under-perform. We recommend Wu et al. for further reading about the capacity for self-detection of specific models. 

\subsection{Academic Integrity in Computing Education}
Research focused on the intersection computing education with LLMs has begun to emerge over the last few years. The work of \textcite{zhu2024embracingaieducationunderstanding} concentrates on secondary education students, and many of these will matriculate into post-secondary education environments. For our purposes it is the technical nature of computing content that presents additional detection challenges. Programming concepts, algorithm explanations, and documentation often follow  patterns that may resemble LLM output even when written by humans. Correct detection of both human and LLM content is non-trivial in academic settings and tends to be biased towards identifying inputs as human \parencite{Weber-Wulff2023}. Human evaluators (who were academic staff) also appear to struggle with detecting LLM written content, performing worse than AI detection tools \parencite{GPT4AcademicDetection}. This work also noticed that the mean grade of the AI generated work was roughly equivalent to the student mean grade, demonstrating that the capacity of LLMs to solve fundamental problems in non-computing domains where more subjectivity is required is now substantial. For computing education, this lack of subjectivity may harm detection efforts as the correctness of technical content can be objectively verified, potentially obfuscating the distinction between human and LLM as there is generally a `correct' answer (or collection of functionally equivalent solutions) to a problem.

\noindent\subsection{Current Gaps and Research Opportunities}
While existing research has made significant strides in understanding LLM detection capabilities, there are several gaps  especially relevant to educational applications:
\\

\noindent\textbf{Limited Domain-Specific Analysis}: Most detection studies focus on general text domains, with limited attention to specialized academic fields like computing education where technical vocabulary and structured explanations are common. A notable exception is that of LLM generated code detection, discussed previously. However, programming is only a subset of computing education, many aspects borrow from other aspects of computer science and domains like mathematics, statistics, systems engineering, etc.
\\

\noindent\textbf{Limited Self-Detection Research:} The capacity for LLMs to detect their own (or other related model) outputs remains underexplored. For educational contexts where understanding model capabilities that may be used to make academic misconduct decisions this is crucial for policy development.
\\

\noindent\textbf{Lack of Adversarial Evaluation}: Limited research has examined detection performance when LLMs are explicitly instructed to evade detection, a scenario increasingly relevant as both the models become more sophisticated and users become more familiar with efficient prompt creation.
\\

Our research addresses these gaps by evaluating LLM self-detection capabilities with respect to common topics within computing education, examining both standard and adversarial conditions. This work contributes to the growing body of literature on AI detection while specifically addressing the practical needs of computing educators and provides a basis for educational policy development.

\section{Methodology}
To investigate the self-detection capabilities of Large Language Models (LLMs) in a computing education context, we designed and executed an experiment to test the capacity for LLM self-reflection by collecting data from a series of prompts that ask for a solution or explanation to some question. These questions span common topics in computer science. 

\subsection{Research Questions}
This experiment was designed to answer the following research questions, which are motivated by the need for reliable AI detection methods in academic settings:
\begin{enumerate}
    \item How accurately do prominent LLMs (GPT-4, Claude, Gemini) classify text as either AI-generated or human-written within the domain of computing education?
    \item What is the false-positive rate of these LLMs when evaluating verified human-authored academic text, and how reliable are their self-reported confidence scores? 
    \item How is detection accuracy affected when a generating LLM is given an adversarial prompt explicitly instructing it to evade detection? 
\end{enumerate}
\subsection{Variables}
\textbf{Independent Variables:}
\begin{itemize}
    \item \textbf{Text Source}: The origin of the text being evaluated. This categorical variable included eight levels: Human-authored, GPT-4 (default), GPT-4 (deceptive), Claude (default), Claude (deceptive), Gemini (default), and Gemini (deceptive).
    \item \textbf{Evaluating LLM}: The model performing the classification task (GPT-4 Turbo, Claude 4 Sonnet, Gemini 2.5 Pro).
\end{itemize}
\textbf{Dependent Variables:}
\begin{itemize}
    \item \textbf{Classification}: The binary label assigned by the evaluating LLM: "AI-generated" or "Human-written".
    \item \textbf{Reported Confidence}: The numerical value (0-100\%) of certainty reported by the model for its classification.
    \item \textbf{Qualitative Reasoning}: The textual justification provided by the model for its decision. Note that this was collected (and is made available in the raw data provided) but ultimately not used in the analysis.
\end{itemize}
\subsection{Materials and Corpus Development}
\noindent \textbf{LLM Selection}: We selected three state-of-the-art, publicly accessible LLMs from distinct developers: OpenAI's GPT-4 Turbo, Google's Gemini 2.5 Pro, and Anthropic's Claude 4 Sonnet. This selection represents a sample of the most popular and powerful models available to students. These models were chosen not only for their popularity but also because they represent distinct underlying architectures and training philosophies, providing a more robust test of detection capabilities across a variety of systems. We were also interested in assessing the current self-detection capabilities of the Claude model family, as prior work had demonstrated no such capacity in earlier versions \parencite{caiado2023aicontentselfdetectiontransformerbased}. 
\\

\noindent \textbf{Model and Environment Details}:All data for this study was collected manually using the freely available public web interfaces for each large language model. No APIs were used. All generation and evaluation sessions were conducted in May and early June of 2025. The specific models and access paths are as follows:
\begin{itemize}
    \item \textbf{GPT-4 Turbo}: Accessed via the ChatGPT web interface at chat.openai.com. The version used was the standard GPT-4 Turbo model available to free-tier users during the data collection period.
    \item \textbf{Gemini 2.5 Pro}: Accessed via the Gemini web interface at gemini.google.com. This version used was the premium version available to paid users during the data collection period.
    \item \textbf{Claude 4 Sonnet}: Accessed via the Claude web interface at claude.ai. The version used was the standard Sonnet 4 model available to free-tier users during the data collection period.
\end{itemize}

As the public web interfaces were utilized, the default, non-configurable parameters for each service were employed for all interactions. This includes parameters such as temperature, top\_p, and maximum token counts, which are not user-adjustable in these environments. Similarly, the default system prompts and configurations inherent to each service were used without modification. Each query was processed in a new, distinct chat session to prevent conversational history from influencing the outputs.
\\

\noindent \textbf{Prompt Design}: We designed five prompt categories to mirror a representative cross-section of common assessment tasks in undergraduate computer science curricula. The categories covered are: 
\\

\noindent\textbf{1 - Algorithm explanations: }`\textit{Explain how binary search works, and what conditions are necessary to use it.}'

\noindent\textbf{2 - Descriptions of technical concepts}: `\textit{What is virtual memory?}'

\noindent\textbf{3 - Documentation of code}: `\textit{Given the following function, write documentation for its purpose and the major components.}'

\noindent\textbf{4 - System design}: `\textit{Outline a simple database system.}'

\noindent\textbf{5 - General problem solving}: `\textit{Design a hash function that satisfies the following conditions: ...}'
\\

\noindent \textbf{Human Baseline Corpus}: A primary concern was evaluating the false-positive rate, which is a critical factor for academic integrity applications where false accusations are a significantly worse outcome than missed detections. To do so we constructed a baseline corpus of verified human-authored text. This corpus was compiled from textbooks, research papers, and personal classroom materials guaranteed to be authored before 2020, predating the widespread availability of powerful LLMs.

\subsection{Experimental Procedure}
\noindent The experiment was conducted in two phases: data generation and evaluation.
\\
\noindent \textbf{Data Generation:}

\begin{itemize}
    \item \textbf{Standard Responses}: For each of the five prompt categories, a standard response was generated from each of the three selected LLMs.
    \item \textbf{Deceptive Responses}: To test the models' resilience to adversarial manipulation, we generated a second set of responses. For this set, the original prompts were prefixed with an instruction for the LLM to write in a style that would not be easily identified as AI-generated, asking it to vary sentence structure, use colloquialisms, and avoid formal patterns.
\end{itemize}
\noindent \textbf{Evaluation:}

\begin{itemize}
    \item Each of the generated texts and each of the human-authored baseline texts were fed into each of the three evaluating LLMs.
    \item To prevent data from prior inputs tainting the results, every evaluation was conducted in a new and distinct session.
    \item For each evaluation, the LLM was given the following standardized prompt:
\end{itemize}
\textit{\begin{quote}
Please analyze the following text and determine if it was written by an AI language model or by a human. Provide your assessment as:
\\\\
1.	Classification: AI-generated or Human-written\\
2.	Confidence level: 0-100\% (how certain are you?)\\
3.	Brief reasoning: What features led to your decision?\\
\\
Text to analyze: [TEXT]
\end{quote}}

\begin{table*}[t]
    \caption{Human Baseline Performance}
    \centering
     \footnotesize
    \begin{tabular*}{\textwidth}{c@{\hspace{3pt}}c@{\extracolsep{\fill}}ccccccc}
        \toprule
        & & \multicolumn{2}{c}{\textbf{GPT-4 Turbo}} & \multicolumn{2}{c}{\textbf{Claude 4 Sonnet}} & \multicolumn{2}{c}{\textbf{Gemini 2.5 Pro}} \\
        \cmidrule(lr){3-4} \cmidrule(lr){5-6} \cmidrule(lr){7-8}
        & \textbf{Prompt} & \textbf{Class.} & \textbf{Conf. (\%)} & \textbf{Class.} & \textbf{Conf. (\%)} & \textbf{Class.} & \textbf{Conf. (\%)} \\
        \midrule
        \multirow{5}{*}{\rotatebox[origin=c]{90}{\textbf{Category 1}}} 
        & 1 & Human & 95\% & Human & 85\% & Human & 90\% \\
        & 2 & AI & 85\% & Human & 75\% & Human & 70\% \\
        & 3 & AI & 95\% & Human & 85\% & Human & 85\% \\
        & 4 & AI & 95\% & Human & 85\% & Human & 90\% \\
        & 5 & AI & 95\% & AI & 85\% & AI & 85\% \\
        \midrule
        \multirow{5}{*}{\rotatebox[origin=c]{90}{\textbf{Category 2}}} 
        & 1 & AI & 95\% & Human & 75\% & Human & 85\% \\
        & 2 & Human & 95\% & Human & 85\% & Human & 90\% \\
        & 3 & AI & 95\% & Human & 75\% & Human & 80\% \\
        & 4 & AI & 85\% & Human & 85\% & Human & 80\% \\
        & 5 & AI & 90\% & Human & 85\% & Human & 80\% \\
        \midrule
        \multirow{5}{*}{\rotatebox[origin=c]{90}{\textbf{Category 3}}} 
        & 1 & AI & 90\% & Human & 75\% & Human & 85\% \\
        & 2 & AI & 95\% & Human & 75\% & AI & 90\% \\
        & 3 & AI & 90\% & Human & 75\% & Human & 85\% \\
        & 4 & AI & 85\% & AI & 75\% & AI & 90\% \\
        & 5 & AI & 90\% & Human & 75\% & AI & 95\% \\
        \midrule
        \multirow{5}{*}{\rotatebox[origin=c]{90}{\textbf{Category 4}}} 
        & 1 & AI & 95\% & AI & 85\% & Human & 85\% \\
        & 2 & AI & 95\% & AI & 85\% & AI & 85\% \\
        & 3 & AI & 95\% & AI & 85\% & AI & 85\% \\
        & 4 & AI & 95\% & Human & 75\% & Human & 80\% \\
        & 5 & AI & 85\% & Human & 75\% & Human & 85\% \\
        \midrule
        \multirow{5}{*}{\rotatebox[origin=c]{90}{\textbf{Category 5}}} 
        & 1 & Human & 95\% & Human & 85\% & Human & 95\% \\
        & 2 & AI & 95\% & AI & 78\% & Human & 90\% \\
        & 3 & AI & 95\% & Human & 75\% & AI & 85\% \\
        & 4 & AI & 90\% & Human & 75\% & Human & 75\% \\
        & 5 & AI & 95\% & AI & 85\% & AI & 90\% \\
        \midrule
        \multicolumn{2}{l}{\textbf{Accuracy}} & 3 / 25 (12\%) &  & 18 / 25 (72\%) &  &  17 / 25 (68\%)&  \\
        \bottomrule
    \end{tabular*}
    
    \label{tab:baseline_raw_rotated}
\end{table*}

\noindent \textbf{Incorporating Deception:} Even if models possess substantial self-detection capabilities (we include cross-model detection as well) on standard output, if simple changes to the response suffice to break self-detection then the usage of models for LLM detection is still unable to be justified.
\\

\noindent The adjustment to the standard prompt is as follows:
\textit{\begin{quote}
    Please respond to the following prompt while deliberately writing in a style that would not be easily identified as AI-generated. Vary your sentence structure, include minor imperfections or colloquialisms that humans might use, and avoid overly structured or formal language patterns. Especially avoid the Title1: Response Response Response Title2: Response Response … structure and format. \\\\ Here is the prompt: [PROMPT]
\end{quote}}

\noindent The final sentence before the prompt was added after initial testing showing that this familiar LLM style structure (see nearly any standard ChatGPT response to a query asking for an explanation or summary) was common even if the underlying content was changed.

\subsection{Data Analysis:} 
Our analysis focused on descriptive statistics to evaluate model performance. For each condition, we calculated the classification accuracy as a percentage and reported it with raw counts. For each evaluating LLM, we calculated its accuracy in classifying texts from all eight source conditions. The error rate was calculated for the human baseline to determine the false-positive rate. We also analyzed the self-reported confidence scores to determine if a correlation existed between a model's stated confidence and its actual accuracy.

\begin{table*}[t]

        \caption{Cross-Model Evaluation Results on LLM Generated Inputs}
    \centering
    \footnotesize 
    \setlength{\tabcolsep}{3pt} 

    \begin{tabular*}{\textwidth}{c@{\hspace{3pt}}c@{\extracolsep{\fill}}cccccc|cccccc|cccccc}
        \toprule
        & & \multicolumn{6}{c}{\textbf{Results from GPT-4}} & \multicolumn{6}{c}{\textbf{Results from Claude}} & \multicolumn{6}{c}{\textbf{Results from Gemini}} \\
        \cmidrule(lr){3-8} \cmidrule(lr){9-14} \cmidrule(lr){15-20}

        & \begin{tabular}{@{}c@{}}\textbf{Source:}\end{tabular} & \multicolumn{2}{c}{\textbf{GPT-4}} & \multicolumn{2}{c}{\textbf{Claude}} & \multicolumn{2}{c}{\textbf{Gemini}} & \multicolumn{2}{c}{\textbf{GPT-4}} & \multicolumn{2}{c}{\textbf{Claude}} & \multicolumn{2}{c}{\textbf{Gemini}} & \multicolumn{2}{c}{\textbf{GPT-4}} & \multicolumn{2}{c}{\textbf{Claude}} & \multicolumn{2}{c}{\textbf{Gemini}} \\
        \cmidrule(lr){3-4} \cmidrule(lr){5-6} \cmidrule(lr){7-8} \cmidrule(lr){9-10} \cmidrule(lr){11-12} \cmidrule(lr){13-14} \cmidrule(lr){15-16} \cmidrule(lr){17-18} \cmidrule(lr){19-20}

        & \textbf{\#} & Cls. & C\% & Cls. & C\% & Cls. & C\% & Cls. & C\% & Cls. & C\% & Cls. & C\% & Cls. & C\% & Cls. & C\% & Cls. & C\% \\
        \midrule
        \multirow{5}{*}{\rotatebox[origin=c]{90}{\textbf{Category 1}}} 
        & 1 & AI & 95\% & AI & 85\% & AI & 90\%  & AI & 95\% & AI & 85\% & AI & 90\% & AI & 95\% & AI & 85\% & AI & 95\% \\
        & 2 & AI & 95\% & AI & 85\% & AI & 90\% & AI & 95\% & AI & 85\% & AI & 90\% & AI & 95\% & AI & 85\% & AI & 95\% \\
        & 3 & AI & 95\% & AI & 85\% & AI & 90\% & AI & 95\% & AI & 85\% & AI & 95\% & AI & 95\% & AI & 85\% & AI & 95\% \\
        & 4 & AI & 95\% & AI & 78\% & AI & 90\% & AI & 95\% & AI & 85\% & AI & 90\% & AI & 95\% & AI & 85\% & AI & 95\% \\
        & 5 & AI & 95\% & AI & 85\% & AI & 90\% & AI & 98\% & AI & 85\% & AI & 90\% & AI & 95\% & AI & 85\% & AI & 95\% \\
        \midrule
        \multirow{5}{*}{\rotatebox[origin=c]{90}{\textbf{Category 2}}} 
        & 1 & AI & 95\% & AI & 85\% & AI & 90\% & AI & 95\% & AI & 85\% & AI & 85\% & AI & 95\% & AI & 85\% & AI & 95\% \\
        & 2 & AI & 95\% & H & 85\% & AI & 90\% & AI & 95\% & H & 75\% & AI & 90\% & AI & 95\% & AI & 85\% & AI & 95\% \\
        & 3 & AI & 95\% & AI & 85\% & AI & 90\% & AI & 95\% & AI & 85\% & AI & 85\% & AI & 95\% & AI & 85\% & AI & 98\% \\
        & 4 & AI & 95\% & AI & 75\% & H & 70\% & AI & 95\% & AI & 85\% & H & 75\% & AI & 95\% & AI & 85\% & AI & 95\% \\
        & 5 & AI & 95\% & AI & 85\% & AI & 85\% & AI & 95\% & AI & 85\% & H & 85\% & AI & 95\% & AI & 85\% & AI & 98\% \\
        \midrule
        \multirow{5}{*}{\rotatebox[origin=c]{90}{\textbf{Category 3}}} 
        & 1 & AI & 95\% & AI & 85\% & AI & 90\% & AI & 95\% & AI & 95\% & AI & 95\% & AI & 95\% & AI & 85\% & AI & 95\% \\
        & 2 & AI & 95\% & AI & 85\% & AI & 90\% & AI & 99\% & AI & 95\% & AI & 95\% & AI & 95\% & AI & 85\% & AI & 95\% \\
        & 3 & AI & 95\% & AI & 85\% & AI & 90\% & AI & 98\% & AI & 88\% & AI & 95\% & AI & 98\% & AI & 85\% & AI & 95\% \\
        & 4 & AI & 95\% & AI & 85\% & AI & 90\% & AI & 98\% & AI & 95\% & AI & 95\% & AI & 98\% & AI & 85\% & AI & 95\% \\
        & 5 & AI & 85\% & AI & 75\% & AI & 95\% & AI & 98\% & AI & 85\% & AI & 95\% & AI & 95\% & AI & 85\% & AI & 95\% \\
        \midrule
        \multirow{5}{*}{\rotatebox[origin=c]{90}{\textbf{Category 4}}} 
        & 1 & AI & 85\% & AI & 85\% & AI & 95\% & AI & 95\% & AI & 95\% & AI & 95\% & AI & 95\% & AI & 88\% & AI & 98\% \\
        & 2 & AI & 85\% & AI & 85\% & AI & 90\% & AI & 95\% & AI & 95\% & AI & 95\% & AI & 95\% & AI & 85\% & AI & 95\% \\
        & 3 & AI & 85\% & AI & 85\% & AI & 95\% & AI & 99\% & AI & 88\% & AI & 95\% & AI & 98\% & AI & 85\% & AI & 98\% \\
        & 4 & AI & 85\% & AI & 95\% & AI & 95\% & AI & 99\% & AI & 88\% & AI & 95\% & AI & 98\% & AI & 88\% & AI & 98\% \\
        & 5 & AI & 85\% & AI & 85\% & AI & 90\% & AI & 98\% & AI & 93\% & AI & 95\% & AI & 95\% & AI & 85\% & AI & 98\% \\
        \midrule
        \multirow{5}{*}{\rotatebox[origin=c]{90}{\textbf{Category 5}}} 
        & 1 & AI & 85\% & AI & 85\% & AI & 95\% & AI & 99\% & AI & 85\% & AI & 95\% & AI & 95\% & AI & 85\% & AI & 95\% \\
        & 2 & AI & 85\% & AI & 85\% & AI & 95\% & AI & 98\% & AI & 87\% & AI & 90\% & AI & 95\% & AI & 85\% & AI & 95\% \\
        & 3 & AI & 85\% & AI & 85\% & AI & 90\% & AI & 95\% & AI & 85\% & AI & 95\% & AI & 98\% & AI & 88\% & AI & 95\% \\
        & 4 & AI & 85\% & AI & 85\% & AI & 90\% & AI & 98\% & AI & 87\% & AI & 95\% & AI & 98\% & AI & 88\% & AI & 95\% \\
        & 5 & AI & 85\% & AI & 85\% & AI & 95\% & AI & 98\% & AI & 85\% & AI & 95\% & AI & 95\% & AI & 85\% & AI & 95\% \\
        \midrule
        \multicolumn{2}{l}{\textbf{Accuracy}} & 25 / 25 &  & 24 / 25 &  & 24 / 25 & &  25 / 25 &  & 24 / 25 & & 23 / 25 &  & 25 / 25 & & 25 / 25 &  & 25 / 25 &\\
        \multicolumn{2}{l}{\textbf{}} & (100\%) &  & (96\%) &  & (96\%) & &  (100\%) &  & (96\%) & & (92\%) &  & (100\%) & & (100\%) &  & (100\%) &\\
        \bottomrule
    \end{tabular*}

    \label{tab:cross_model_results_default}

    \begin{tablenotes}
            \item[*] \textit{Note: Cls is classification with results of AI or Human (H). C\% is the model reported confidence in its prediction.}
        \end{tablenotes}

\end{table*}

\begin{table*}[t]
        \caption{Cross-Model Evaluation Results on Deceptive LLM Generated Inputs}
    \centering
    \footnotesize 
    \setlength{\tabcolsep}{3pt} 

    \begin{tabular*}{\textwidth}{c@{\hspace{3pt}}c@{\extracolsep{\fill}}cccccc|cccccc|cccccc}
        \toprule
        & & \multicolumn{6}{c}{\textbf{Results from GPT-4}} & \multicolumn{6}{c}{\textbf{Results from Claude}} & \multicolumn{6}{c}{\textbf{Results from Gemini}} \\
        \cmidrule(lr){3-8} \cmidrule(lr){9-14} \cmidrule(lr){15-20}

        & \begin{tabular}{@{}c@{}}\textbf{Source:}\end{tabular} & \multicolumn{2}{c}{\textbf{GPT-4}} & \multicolumn{2}{c}{\textbf{Claude}} & \multicolumn{2}{c}{\textbf{Gemini}} & \multicolumn{2}{c}{\textbf{GPT-4}} & \multicolumn{2}{c}{\textbf{Claude}} & \multicolumn{2}{c}{\textbf{Gemini}} & \multicolumn{2}{c}{\textbf{GPT-4}} & \multicolumn{2}{c}{\textbf{Claude}} & \multicolumn{2}{c}{\textbf{Gemini}} \\
        \cmidrule(lr){3-4} \cmidrule(lr){5-6} \cmidrule(lr){7-8} \cmidrule(lr){9-10} \cmidrule(lr){11-12} \cmidrule(lr){13-14} \cmidrule(lr){15-16} \cmidrule(lr){17-18} \cmidrule(lr){19-20}

        & \textbf{\#} & Cls. & C\% & Cls. & C\% & Cls. & C\% & Cls. & C\% & Cls. & C\% & Cls. & C\% & Cls. & C\% & Cls. & C\% & Cls. & C\% \\
        \midrule
        \multirow{5}{*}{\rotatebox[origin=c]{90}{\textbf{Category 1}}} 
        & 1 & H & 85\% & AI & 95\% & H & 90\% & AI & 95\% & H & 85\% & H & 90\% & AI & 95\% & AI & 85\% & H & 85\% \\
        & 2 & H & 85\% & AI & 95\% & H & 85\% & H & 85\% & H & 85\% & H & 85\% & H & 85\% & H & 85\% & H & 85\% \\
        & 3 & H & 85\% & AI & 95\% & H & 85\% & H & 90\% & H & 85\% & H & 90\% & H & 90\% & H & 85\% & H & 90\% \\
        & 4 & AI & 85\% & AI & 95\% & H & 85\% & H & 90\% & H & 85\% & H & 85\% & H & 90\% & H & 85\% & H & 85\% \\
        & 5 & AI & 95\% & AI & 95\% & H & 85\% & AI & 95\% & H & 85\% & H & 85\% & AI & 95\% & H & 85\% & H & 85\% \\
        \midrule
        \multirow{5}{*}{\rotatebox[origin=c]{90}{\textbf{Category 2}}} 
        & 1 & AI & 92\% & AI & 95\% & H & 90\% & H & 90\% & H & 85\% & H & 90\% & H & 90\% & H & 85\% & H & 90\% \\
        & 2 & AI & 95\% & AI & 95\% & H & 85\% & AI & 95\% & H & 85\% & H & 85\% & AI & 95\% & H & 85\% & H & 85\% \\
        & 3 & AI & 95\% & AI & 95\% & H & 85\% & AI & 95\% & H & 85\% & H & 90\% & AI & 95\% & AI & 85\% & AI & 90\% \\
        & 4 & AI & 95\% & AI & 95\% & H & 90\% & AI & 95\% & H & 85\% & H & 90\% & AI & 95\% & H & 85\% & H & 90\% \\
        & 5 & AI & 95\% & AI & 95\% & H & 85\% & AI & 95\% & H & 85\% & H & 85\% & AI & 95\% & H & 85\% & H & 85\% \\
        \midrule
        \multirow{5}{*}{\rotatebox[origin=c]{90}{\textbf{Category 3}}} 
        & 1 & AI & 95\% & AI & 95\% & H & 90\% & AI & 95\% & AI & 85\% & AI & 95\% & AI & 95\% & H & 85\% & H & 85\% \\
        & 2 & AI & 95\% & AI & 95\% & H & 90\% & AI & 95\% & H & 85\% & H & 85\% & AI & 95\% & AI & 85\% & H & 90\% \\
        & 3 & AI & 90\% & AI & 95\% & H & 90\% & AI & 95\% & H & 85\% & AI & 95\% & AI & 95\% & H & 85\% & H & 90\% \\
        & 4 & AI & 95\% & AI & 95\% & H & 90\% & AI & 95\% & H & 85\% & H & 85\% & AI & 95\% & H & 85\% & H & 90\% \\
        & 5 & AI & 95\% & AI & 95\% & H & 90\% & AI & 95\% & H & 85\% & H & 85\% & AI & 95\% & H & 85\% & H & 85\% \\
        \midrule
        \multirow{5}{*}{\rotatebox[origin=c]{90}{\textbf{Category 4}}} 
        & 1 & AI & 95\% & AI & 95\% & H & 90\% & AI & 95\% & AI & 85\% & H & 90\% & AI & 95\% & H & 75\% & H & 90\% \\
        & 2 & AI & 95\% & AI & 95\% & H & 85\% & AI & 95\% & H & 85\% & H & 85\% & AI & 95\% & H & 85\% & H & 90\% \\
        & 3 & AI & 95\% & AI & 95\% & H & 90\% & AI & 95\% & H & 85\% & H & 85\% & AI & 95\% & H & 85\% & H & 90\% \\
        & 4 & AI & 95\% & AI & 95\% & H & 85\% & AI & 95\% & AI & 85\% & H & 90\% & AI & 95\% & H & 85\% & H & 90\% \\
        & 5 & AI & 95\% & AI & 95\% & H & 85\% & AI & 95\% & AI & 85\% & AI & 95\% & AI & 95\% & H & 85\% & H & 90\% \\
        \midrule
        \multirow{5}{*}{\rotatebox[origin=c]{90}{\textbf{Category 5}}} 
        & 1 & AI & 95\% & AI & 95\% & H & 85\% & AI & 95\% & H & 85\% & AI & 90\% & AI & 95\% & AI & 85\% & H & 85\% \\
        & 2 & AI & 92\% & AI & 95\% & H & 85\% & AI & 95\% & H & 85\% & H & 90\% & AI & 95\% & H & 85\% & H & 85\% \\
        & 3 & AI & 96\% & AI & 95\% & H & 85\% & AI & 95\% & H & 85\% & H & 85\% & AI & 95\% & H & 85\% & H & 85\% \\
        & 4 & AI & 95\% & AI & 95\% & H & 90\% & AI & 95\% & H & 85\% & H & 90\% & AI & 95\% & H & 85\% & H & 90\% \\
        & 5 & AI & 90\% & AI & 95\% & H & 90\% & AI & 95\% & H & 85\% & H & 90\% & AI & 95\% & H & 85\% & H & 95\% \\
        \midrule
        \multicolumn{2}{l}{\textbf{Accuracy}} & 22 / 25 &  & 25 / 25 &  & 0 / 25 & &  21 / 25 &  & 4 / 25 & & 4 / 25 &  & 21 / 25 & & 4 / 25 &  & 1 / 25 & \\
        \multicolumn{2}{l}{\textbf{}} & (88\%) &  & (100\%) &  & (0\%) & &  (84\%) &  & (16\%) & & (16\%) &  & (84\%) & & (16\%) &  & (4\%) &\\
        \bottomrule
        
    \end{tabular*}
        
    \label{tab:cross_model_results_deceptive}

    \begin{tablenotes}
            \item[*] \textit{Note: Cls is classification with results of AI or Human (H). C\% is the model reported confidence in its prediction.}
        \end{tablenotes}
\end{table*}

\section{Results and Discussion}
\subsection{Baseline}\label{sec:baseline}
The baseline results (Table \ref{tab:baseline_raw_rotated}) show poor performance for every model. GPT-4 was noticeably biased toward predicting nearly all of the human inputs as AI (88\%). While Claude and Gemini performed substantially better, their error rate was appreciable at 28\% and 32\% respectively, too high to consider as a reliable source of information for determining plagiarism. 
The self-reported confidence scores provided by the models appear to have no diagnostic value. We found no meaningful correlation between the stated confidence level and the accuracy of the classification. For instance, models would often report high confidence (e.g., 90\%+) for classifications that were incorrect. The overlap in confidence ranges for correct and incorrect judgments demonstrates the unreliability of this metric for determining accuracy (Table \ref{tab:classifications}). Ultimately, the reported confidence is unreliable and that the textual justifications accompanying them appear to be spurious post-hoc rationalizations rather than genuine reasoning.
\subsection{LLM Self-Detection}
\noindent \textbf{Default LLM Responses:} The default LLM generated outputs were overwhelmingly classified correctly (Table \ref{tab:cross_model_results_default}). The worst performance (Claude classifying Gemini's output) was still 92\% accurate. Gemini's performance was flawless, with zero misclassifications. GPT-4 performed well (96\% on Claude and Gemini outputs), with  but as seen with the human baseline it may be that the model is biased towards predicting a classification of AI written (the deceptive analysis below does provide evidence against this conjecture). Gemini and GPT-4 displayed perfect performance at identifying their own output, while Claude suffered from one error (96\% accurate). The Gemini output was the most effective at producing misclassifications with one for GPT-4 and two for Claude. Claude's output produced two misclassifications, one for GPT-4 and one for itself. GPT-4's output was always detected correctly, which is not surprising as it follows a distinctive convention of how it arranges its output.

Again, the confidence levels appear to be of no real use for the reasons mentioned in Section \ref{sec:baseline}. We do note that Claude's misclassifications produced lower confidence values for the misclassified examples in comparison to the most common values for the remainder.
\\

\noindent \textbf{Deceptive LLM Responses:} The deceptive prompts showed substantial effectiveness on certain models (Table \ref{tab:cross_model_results_deceptive}). 
In sharp contrast to the default condition, the use of deceptive prompts caused a dramatic degradation in detection accuracy across all models (Table \ref{tab:cross_model_results_deceptive}). 
The effect was most pronounced on deceptive text sourced from Gemini. GPT-4’s detection accuracy for this text collapsed to 0\% (0/25 correct). Claude’s performance was also severely impacted, with its accuracy falling to 16\% (4/25 correct). Even more strikingly, Gemini was able to deceive itself 96\% of the time (24/25 incorrect), highlighting a substantial incapacity.


A key qualitative difference we observed was that all models adopted a more informal, conversational tone when prompted for deception, frequently using contractions and colloquialisms (Table \ref{tab:my_experiment_results}). This stylistic shift appears to be an important factor in evading detection, likely because it breaks the statistical patterns associated with the models' default mode of generation (which is more formal). While from a human perspective the explanations were still clear, this tonal shift was sufficient to fool the classifiers, suggesting their detection algorithms may be brittle and overly reliant on stylistic markers.

As all models were vulnerable to some level at what are effectively trivial changes to the prompt structure in detecting LLM output, these models should not be used as the basis for judgment in academic misconduct cases. Overall, the output of these models is too unstable to be trusted, as these simple alterations to content and structure significantly degrade the efficacy of these models.

\begin{table}[ht] \label{tab:classifications}
\centering
\footnotesize
\caption{Summary statistics for model-reported confidence scores (\%). Scores are grouped by the evaluating model and the classification's correctness.}
\label{tab:confidence_stats}
\begin{tabular}{@{}lcccc@{}}
\toprule
\multicolumn{5}{c}{\textbf{Correct Classification}} \\ \cmidrule(lr){1-5}
\textbf{Evaluating Model} & \textbf{Mean} & \textbf{Min} & \textbf{Max} & \textbf{Mode} \\ \midrule
\textbf{GPT-4} & 89.9 & 75 & 96 & 95 \\
\textbf{Gemini} & 87.8 & 70 & 98 & 95 \\
\textbf{Claude} & 87.5 & 75 & 99 & 85 \\ \midrule
\multicolumn{5}{c}{\textbf{Incorrect Classification}} \\ \cmidrule(lr){1-5}
\textbf{Evaluating Model} & \textbf{Mean} & \textbf{Min} & \textbf{Max} & \textbf{Mode} \\ \midrule
\textbf{GPT-4} & 89.1 & 85 & 95 & 85 \\
\textbf{Gemini} & 87.9 & 75 & 95 & 90 \\
\textbf{Claude} & 83.5 & 75 & 95 & 85 \\ \bottomrule
\end{tabular}
\end{table}

\begin{table*}[ht]
    \caption{Examples of Default and Deceptive Generated Prompts}
    \centering
    \footnotesize 

    \begin{tabularx}{\textwidth}{>{\RaggedRight}X | >{\RaggedRight}X | >{\RaggedRight\arraybackslash}X}
        \toprule
        \multicolumn{3}{c}{\textbf{Prompt}} \\
        \multicolumn{3}{c}{ `Explain how the binary search algorithm works and why it is more efficient than linear search.'} \\
        \cmidrule(lr){1-3}
        
        \textbf{GPT-4} & \textbf{Claude} & \textbf{Gemini} \\
        \midrule
        
        \multicolumn{3}{c}{\textbf{Default}} \\
        \cmidrule(lr){1-3}
        Binary search is an efficient algorithm used to find the position of a target element within a sorted list or array. It works by repeatedly dividing the search interval in half$\dots$
        & 
        Binary search is a highly efficient algorithm for finding a specific value in a sorted array or list. It works by repeatedly dividing the search space in half until the target element is found or determined to be absent$\dots$
        & 
        The binary search algorithm is a highly efficient search method used to find a specific element within a sorted array or list. Its efficiency stems from a "divide and conquer" strategy, which dramatically reduces the number of comparisons needed compared to a linear search$\dots$
        \\
        \midrule

        \multicolumn{3}{c}{\textbf{Deceptive}} \\
        \cmidrule(lr){1-3}
        Alright, so imagine you're looking for a name in a really big phone book—like, one of those old-school ones, not digital. If you start at the beginning and check each name one by one until you find what you’re looking for, that’s basically how linear search works. It doesn’t matter if the list is sorted or not; you just keep going line by line. Simple, but slow if the list is huge.
Now, binary search is like the smarter cousin. It only works if the list is already sorted (which is key)$\dots$
 & 
        Alright, so binary search - it's actually pretty clever when you think about it. You know how when you're looking for a word in the dictionary, you don't start from page 1 and flip through every single page? You kinda jump to the middle and see if you've gone too far or not far enough, then adjust from there. That's basically what binary search does, but with data.
Here's the thing though - and this is important - your data has to be sorted first$\dots$
& 
        Alright, let's break down this binary search thing, shall we? You know, how it actually gets the job done and why it's such a superstar compared to just trudging along with a linear search.
So, imagine you've got a phone book, and it's a massive one. If you wanted to find "Smith," you wouldn't start at "Aaronson" and just flip page by page until you hit "Smith," right? That's basically what a linear search does – it just checks every single entry one after another until it finds what it's looking for or runs out of things to check. Takes forever, especially with a big list$\dots$
\\
        
        \bottomrule
    \end{tabularx}
    \label{tab:my_experiment_results}
\end{table*}

\section{Conclusion}

The rapid integration of Large Language Models into academic life has created an urgent need for reliable methods to ensure academic integrity. This paper investigated the viability of using prominent LLMs (GPT-4, Claude Sonnet 4, and Gemini 2.5 Pro) as detectors of AI-generated content. Our findings reveal important limitations in their performance. While the models could easily identify their own default, formulaic outputs, their accuracy largely collapsed using a short and simple shift in the prompt instructions. These findings reinforce the lack of robustness shown by LLMs as classifiers which remains a common vulnerability across deep learning models.

We found that all models performed poorly when evaluating human-authored text, with error rates for some models exceeding 30\%. Importantly, the models proved highly susceptible to simple `deceptive' prompts. When instructed to write in a more informal, human-like tone, the AI-generated text generally fooled not only other models but also the generating model itself. The models' vulnerability to tonal shifts suggests they are engaged in shortcut learning, relying on superficial stylistic markers rather than semantic content to make classifications. This reliance on spurious correlations is what makes them so easily deceived.. Furthermore, our analysis showed that the self-reported confidence scores from these models are effectively useless, bearing no meaningful correlation to accuracy and giving a false sense of trustworthiness. Ultimately, current, publicly accessible LLMs are not reliable enough to be used as arbiters in academic misconduct cases. 

This unreliability means that traditional assignments (e.g., 'explain concept X') are no longer aligned with learning outcomes, as the task can be trivially completed by a non-student (the AI). The failure of technological solutions like self-detection forces a pedagogical conclusion: educators must design assessments that are inherently resistant to AI misuse, such as requiring students to apply concepts to novel scenarios or critique existing work, thereby realigning the assessment with the goal of authentic student learning.

\subsection{Limitations and Future Work}
This study has several limitations that offer avenues for future research. Foremost is the lack of repeated evaluations for an identical input. As our data was collected manually, this proved infeasible. As these models possess a certain level of stochasticity, it is plausible that identical prompts would result in different outputs and so this should be the immediate focus of future work so as to provide a best estimate for the expected failure rate. Our investigation was confined to three specific models at a single point in time; as these technologies evolve, their detection capabilities may change. The deceptive prompts used were also intentionally simple; more sophisticated adversarial attacks could be designed to further probe model vulnerabilities. Finally, our focus was strictly on computing related topics, and the findings may not generalize to other academic domains with different writing conventions.

Future work should expand the scope of models tested and explore a wider range of deceptive strategies. Research could also investigate the development of hybrid detection systems that combine stylistic analysis with other forensic markers. Until such robust and validated tools are available, however, we recommend that universities focus on designing assessments that are inherently resistant to AI misuse.

\section{Data Availability Statement}
To promote transparency and support the reproducibility of this research, all data and materials used in this study are openly available in a public repository. This repository contains the complete, raw text outputs from GPT-4, Claude, and Gemini for both standard and deceptive conditions in CSV format containing the classification judgments, confidence scores, and qualitative reasoning from all evaluating models for every text sample. The repository can be accessed at: {\small \url{https://github.com/christopherburger/llm-self-detection-2026}}

\section{Acknowledgments} \noindent The participation of Karmece Talley and Christina Trotter was supported by the Ronald E. McNair Post-Baccalaureate Achievement Program, U.S. Department of Education (grant \#P217A220247).





\printbibliography

\end{document}